\documentstyle[aps,epsfig,multicol]{revtex}
\begin{document}
\draft
\title { Singularity-free Cosmological Solutions with Non-rotating
Perfect Fluids}
\author{A. K. Raychaudhuri}

\address{Relativity and Cosmological Center, Department of Physics, 
Jadavpur University, 
Kolkata 700032, India.}
\maketitle

\begin{abstract}
The paper establishes the result that solutions of the type described in the
title of the article are only those that have been already presented in the 
literature. The procedure adopted in the paper is somewhat novel - while the 
usual practice is to display an exact solution and then to examine whether it is
singularity free, the present paper discovers the conditions which
a singularity free solution of the desired
type must satistfy. There is no attempt to obtain exact solutions.
Simply, the conditions that were ad-hoc introduced in the deduction of 
singularity free solutions are here shown to follow from the requirement
of non-singularity.
\end{abstract}

\begin{multicols}{2}

\section{Introduction}

A number of singularity theorems proved in the late sixties of the last
century and afterwards led to a belief that a singularity free cosmological
solution in general relativity is not possible unless one violates
the strong energy condition and/or causality. Here we are not intererested 
in a critique of these theorems or of the justification of the belief
about the non-existence of singularity free solutions.
We simply note that when in 1990 Senovilla \cite{seno}
presented a non-singular solution with a perfect fluid having $\rho = 3p > 0$,
it appeared to be something exceptional. Shortly afterwards, Ruiz and 
Senovilla \cite{rseno} presented a family of non-singular solutions for 
perfect fluids with or without an equation of state. None of these 
investigations were motivated by a specific search for non-singular solutions
but came as a by-product of an exhaustive investiagtion of $G_2$ cosmologies 
with the following mathematically simplifying assumptions:

\begin{quote}

(a) The two killing vectors commute and are hyper-surface orthogonal.

(b) The metric tensor components are separable functions i.e., one can
write $g_{\mu\nu} = T_{\mu\nu} R_{\mu\nu}$ where $T^,$s  
and $R^,$s are functions of the time coordinate $t$ and the space coordinate $r$ respectively. 
The coordinates $t$ and $r$ are respectively the lines tangential to 
the velocity vector and the space coordinate orthogonal to the spaces
spanned by the killing vectors.

(c) All the functions $T_{\mu\nu}$ are expressible as powers of a 
single function of $t$.
\end{quote}
Ruiz and Senovilla did not assume cylindrical symmetry
but noted with surprise that all singularity free solutions discovered by them
had this symmetry and wondered whether "this type of symmetry could have some
relevance to the avoidance of singularity."

Somewhat later, Mars \cite{mars} gave up the assumption 
that the killing vectors are hyper-surface orthogonal and used an
"ansatz of separation of variables in comoving coordinates" analogous to the 
assumption (b) above.    
However, the condition of separatibilty as usually understood does not
fully hold good in his solution. [The exception is $g_{\phi\phi}
= r^2 \cosh (2at) + a^2r^4 \cosh^{-1}(2at)$.] 
Mars's solution had the equation of state $\rho =p$. Ruiz and Senovilla have
 clearly demonstrated that the only non-singular solutions satisfying their 
assumptions are given by their equation (58), Mars's work showed that the 
generalization to non-orthogonal $G_2$ adds merely 
one member to the family of non-singular solutions with non-rotating perfect
fluids.

It is not unreasonable to think that these non-singular solutions 
(discovered, so to say accidentally) are just a subset of measure 
zero amongst an infinite family of non-singular solutions - 
the discovery of the particular ones has been facilitated
by their simple mathematical form
originating from the very strong assumptions of Ruiz and Senovilla.
Such a view would have been confirmed if during the last
decade some non-singular solution outside the already discovered 
family were found. But as nothing like that has happened, it 
is tempting to nurse the alternative idea that the already discovered solutions exhaust all possible non-singular solutions with non-rotating
perfect fluids irrespective of any other consideration.
The present investigation was motivated by such an idea and establishes 
this result by deducing all the assumptions of Ruiz and Senovilla as well as Mars from the condition of non-singularity. But first let us spell out
the meaning of different terms used by us:

(i) Non-singularity: It is meant to indicate that none of the
physical or geometrical scalars blow up even at infinity. 
The metric tensor components may blow up (indeed some do blow up in
Ruiz-Senovilla and Mars solutions) at spatial or temporal infinity
but they must be bounded and continuous in the finite region of space-time.
We make no use of the sophisticated idea of geodesic completeness in consideration of
singularity. Rather we feel that identifying singularities with incompleteness
of null and/or timelike geodesics is in the ultimate
analysis 'a piece of opportunism as it allows many theorems to be proved' 
\cite{ear}.

(ii) Non-rotating: The fluid flow is assumed to be hydrodynamic 
so that one can introduce a comoving coordinate system with the velocity 
vector hyper-surface orthogonal, i.e., $\omega^\mu = 
\epsilon^{\mu\nu\alpha\sigma} v_\nu v_{\alpha;\sigma}= 0.$

(iii) Perfect fluid - The energy stress tensor is of the form $T^{\mu}_\nu
=(p + \rho)v^{\mu}v_{\nu}-p\delta^{\mu}_\nu$ in the usual notation.
This means that we are just leaving out viscosity, heat flow etc so that the 
solution presented by Dadhich et al \cite{dadhich} are outside our purview.

(iv) Cosmological: The energy tensor obeys the strong energy condition - in
particular for the perfect fluid we take somewhat stronger constraint 
$ 0 \leq p \leq \rho$. The pressure and the density must not vanish anywhere
except at infinity. We thus exclude the case of a 
complete vacuum or a bounded distribution of matter. We shall exclude
any discontinuity in $T_{\mu\nu}$ - although
that is not inconsistent with absence of singularities, such
discontinuities are unusual for simple cosmological models. This will allow 
us to use a  single coordinate system as globally valid. 

(v) We further introduce the physical assumption that changes of $p$ and $\rho$ are in the same direction for any change of the coordinates. Such a 
condition seems natural for any physically acceptable model. 

It is easy to see that in the absence of rotation, a non-singular
cosmological solution of the desired type must have both acceleration and
expansion non-vanishing. For, in  the absence of acceleration, expansion
will necessarily be associated with a collapse singularity as in the
Friedmann models and the expansion free case is not of interest in 
cosmology. (They are usually studied as stellar models.)

Dadhich \cite{dadhich2} has claimed 
that instead of all the assumptions of Ruiz and Senovilla, if one, besides our 
assumptions,  assumes
that in the comoving coordinate system, the metric tensor is diagonal 
and separable into functions of time and space coordinates, one is uniquely
led to the family discovered by Ruiz and Senovilla. But the additional assumptions
of Dadhich are quite strong and leave out Mars's non-diagonal metric. 
Our treatment is free of these shortcomings and assumes nothing besides the
five conditions stated above.

It may be of some interest to note  a paper by Senovilla \cite{seno2}. 
The Senovilla paper adopts the metric:
$$ ds^2 = T^{2(1+n)} \Sigma ^{2n(n-1)} (-dt^2 + dr^2) +
T^{2(1+n)} \Sigma^{2n} \Sigma'^2 d\phi^2$$
$$ + 
T^{2(1-n)} \Sigma^{2(1-n)}  dz^2 $$
where $T$ is a function of $t$ alone and $\Sigma$ is a function of $r$ alone.
This metric satisfies all the conditions used by Ruiz-Senovilla in their
deduction of the singularity free perfect fluid solutions, which under the circumstances are shown 
to be only such solutions. Specifically the assumptions are existence of
$G_2$, orthogonal separable metric, all time functions
in the metric are powers of a single function. Thus any non-singular solution
(they are not explicitly displayed) that \cite{seno2} may imply which differs from that of the Ruiz-Senovilla will either violate the perfect fluid condition
(note that Senovilla allows $p_r \neq p_z$) or violates the strong energy 
condition. (Senovilla himself points out that there are "almost FLRW model"  
which are singularity free but "violate energy conditions".)

\section {The coordinate system and the functions   $p$, $\rho, 
\dot v_\mu$ and   $\theta$}

The non-rotating condition allows us to write the metric in the form:
\begin{equation}
\label{eq1}
ds^2 = g_{00} dt^2 +g_{ik} dx^i dx^k ~~~(i,k=1,2,3)
\end{equation}
with the fluid velocity components
\begin{equation}
\label{eq2}
v^i = 0, ~~ v^0 = \frac{1}{\sqrt{g_{00}}}
\end{equation}
The acceleration vector has the components
\begin{equation}
\label{eq3}
\dot v_i = -\frac{1}{2}(\ln g_{00}),i  ~~; \dot v_0 = 0.
\end{equation}
Thus the acceleration vector is confined to the three space orthogonal to $v^\mu$and in that space it is a gradient vector. Hence we can align a space-like 
coordinate (say $x^1 = r$) along this vector which
will be orthogonal to the other three coordinate lines and hence (\ref{eq1})
will be reduced to
\begin{equation}
\label{eq4}
ds^2 = g_{00} dt^2 +g_{11} dr^2 + g_{ab} dx^a dx^b
\end{equation}
with
\begin{equation}
\label{eq5}
\dot v_a = 0, ;~~~ (a,b = 2,3)
\end{equation}

We shall consider the domain of the coordinates to extend
from $-\infty $ to $+\infty$. In particular we note
that as the tangent to the $r$ coordinate lines is a gradient
vector, any $r$-line cannot form a closed loop. Thus $r$ has to run from 
$-\infty$ to $+\infty$ (or in case the $r$-lines diverge from a point as the radial 
lines in spherical or cylindrical symmetry, $r$ will run from zero to $\infty$.

One may wonder whether the metric form (\ref{eq4}) is globally
valid with a single coordinate system. However,
we note that our condition (iv) setting our idea of a cosmological 
solution ensures this global validity. Obviously, because of 
(\ref{eq5}) and (\ref{eq3}), $g_{00}$ is a function of $r$ and $t$
only (not of $x^2, x^3$). From the relation $T^{\mu}_{\nu;\mu} = 0$,
we get for perfect fluids,
\begin{equation}
\label{eq6}
p,_i = -\frac{p + \rho}{2} (\ln g_{00}),_i 
\end{equation}
Hence $p$ is also a function of $r$ and $t$ only. Writing
equation (\ref{eq6}) in the form
$$ (p + \rho ) = -\frac{2p,_i}{(\ln  g_{00}),_i}  $$
one sees that $\rho$ is also a function of 
$r$ and $t$ only. 
From the other divergence relation
\begin{equation}
\label{eq7}
 \frac{1}{\sqrt{g_{00}}} \dot  \rho  = -\frac{p +\rho}{2} ~~\theta
\end{equation}
one sees that $\theta$ is also a 
function of 
$r$ and $t$ only. 
Summing up, the scalars $p, \rho, \theta$ and the metric tensor component $g_{00}$
(whose gradient determines the acceleration) are all
functions of 
$r$ and $t$ only. 

\section{The vanishing of some first derivatives}

In two communications \cite{akr1,akr2}, the present author established
that for non-singular, non-rotating cosmological solution, both
the space-time average and the space average (defined in a suitable manner),
of each of the scalars that appear in the Raychaudhuri
equation vanish. These results in turn require 
that for  distributions of perfect fluid which are not bounded, 
these scalars must vanish both at spatial and temporal
infinity. 
Thus $p, \rho,\sigma, \theta, \dot v^{\mu};_{\mu} $ all vanish at $r,t 
\rightarrow \pm \infty$. Consequently the positive definite quantities 
like $p, \rho, \sigma^2,\theta^2$ must have at least one maximum
both in $r$ and $t$.

Consider now the case of pressure $p$. As it is a function of $r$ and $t$, the
maximum in $p$ in $r$ defined by $\frac {\partial p}{\partial r} = 0$ and
$\frac{\partial^2p}{\partial r^2} < 0$ will be a line in the $r,t$ space. Along this
 line we shall have,
\begin{equation}
\label{eq8}
0= d_L(\frac {\partial p}{\partial r}) =  \frac{\partial^2 p}{\partial r^2} dr_L
+ \frac{\partial^2p}{\partial r \partial t} ~~dt_L 
\end{equation}
The subscript $L$ indicating that differentials along the line are to be taken.
Also as $\frac {\partial^2p}{\partial r^2} < 0$ all along the line,
the integral $\int \frac{\partial ^2p}{\partial r^2} dr_L$ will diverge if the
domain of integration be infinite but
$\int \frac{\partial ^2p}{\partial r^2} dr_L$ on integration gives 
$(\frac {\partial p}{\partial r})_L$ and will be finite for all limits of
integration. Thus the only way to satisfy 
(\ref{eq8}) is to have $dr_L = 0$ and $\frac{\partial^2p}{\partial r \partial t}$ 
also zero at all points on the line. 
Thus this line is a $t$-line defined by a constant value of $r$.
	
An exactly similar analysis leads to the conclusion that
$(\frac {\partial p}{\partial t}) = 0$ line is a $t$-constant line, i.e., 
a $r$ coordinate line and over it also  $\frac{\partial^2p}{\partial r \partial t}$
vanishes. choosing the intersection of these two orthogonal lines as the
origin ($r=0, t=0$) we may sum up as follows:
$$\frac {\partial p}{\partial r} =0 ~~~\rm{for}~~ r = 0  $$  
\begin{equation}
\label{eq9}
\frac {\partial p}{\partial t} =0 ~~~\rm{for}~~ t = 0 
\end{equation}
$$ \frac{\partial^2p}{\partial r \partial t} = 0 ~~\rm{if~~ either }~~ r =0 ~~\rm{or}~~ t =0 $$
Exactly similarly,
$$\frac {\partial \rho}{\partial r} =0 ~~~\rm{for}~~ r = ~\rm{constant}~~=r_0 ~~\rm{(say)}  $$   
\begin{equation}
\label{eq10}
\frac {\partial \rho}{\partial t} =0 ~~~\rm{for}~~ t = ~\rm{constant}~~=t_0 ~~\rm{(say)}     
\end{equation}
$$ \frac{\partial^2\rho}{\partial r \partial t} = 0 ~~\rm{if~~ either }~~ r =r_0 ~~\rm{or}~~ t =t_0 $$

We have introduced $r_0,t_0$ to indicte the possibility that the maximum of
$p$ may not coincide with the maximum of $\rho$. However,
Such a situation will go against our assumption (v). Hence we have 
$r_0 = t_0  =0$.

Next consider the case of $(\ln g_{00})$. In view of equation (\ref{eq6}),
the zero of $\frac {\partial p}{\partial r}$ is also a 
zero of $\frac {\partial (\ln g_{00})}{\partial r}$.

From (\ref{eq6}), we get,
$$\int \frac {{\partial p}/\partial r}{p} = -\frac {1}{2}\int (1+\frac {\rho}{p})\frac {\partial (\ln g_{00})}{\partial r} $$
If the integral extends from $-\infty$ to $+\infty$,
$p \rightarrow 0$, and hence as $(1+\frac {\rho}{p})$ is finite and 
positive everywhere, $g_{00} \rightarrow +\infty$ as $ r \rightarrow \pm\infty$.
Again, from (\ref{eq7}), by a suitable similar integration, $ |det ~g_{ik}| \equiv |h| 
\rightarrow +\infty$ as 
$ t \rightarrow \pm\infty$. But $\theta$ and $\dot v^{\mu};_{\mu}$ vanish at the infinities, consequently $g_{00} \rightarrow +\infty$ also at $ t \rightarrow \pm\infty$ and  $|h| 
\rightarrow \infty$ as $ r \rightarrow \pm\infty$. Hence both 
$|h|$ and $g_{00}$ must have minimum with respect to both 
$r$ and $t$.

That these minina will occur at $r=0$ and $t=0$ again follows from (\ref{eq6}),
(\ref{eq7}), (\ref{eq9}) and (\ref{eq10}). A further investigation of the vanishing of $\sigma^2$
and  $\dot v^{\mu};_{\mu}$ at the infinities show that $g_{rr}$ and $\gamma$ 
(the two dimensional metric determinant = $\frac{|h|}{|g_{11}|}$)
also blow up at the infinities.

Thus we get
$$2 \dot v^\mu = \frac {\partial (\ln g_{00})}{\partial r} =0 ~~~\rm{for}~~ r = 0 $$
\begin{equation}
\label{eq11}
\frac {\partial  (\ln g_{00})}{\partial t} =0 ~~~\rm{for}~~ t=0
\end{equation}
$$ \frac{\partial^2 (\ln g_{00})}{\partial r \partial t} = 0 ~~\rm{if~~ either }~~ r =0 ~~\rm{or}~~ t =0$$

Similarly from (\ref{eq7}), we get, 
$$ \frac {\partial \ln |h|}{\partial r} =0 ~~~\rm{for}~~ r = 0 $$
\begin{equation}
\label{eq12}
2 \sqrt {g_{00}} ~~\theta = \frac {\partial  \ln |h|}{\partial t} =0 ~~~\rm{for}~~ t=0
\end{equation}
$$ \frac{\partial^2 \ln h}{\partial r \partial t} = 0 ~~\rm{if~~ either }~~ r =0 ~~\rm{or}~~ t =0$$

The pair of orthogonal lines $r=0$ and $t=0$ are the lines of vanishing 
of first derivatives of $p, \rho, \ln g_{00}, \ln h$.
It is easy to see that at no other value of $r$, can $\frac 
{\partial p}{\partial r}$ vanish. If  possible  let us suppose that at
$r = R(\neq 0), \frac 
{\partial p}{\partial r} =0$.
From our previous analysis, it is clear that non-singularity requires
that this cannot be an isolated point but 
would be the entire $t$ coordinate line at which $r=R$. Also as $r=R$ follows
the maximum of $p$ at $r=0$, $\frac {\partial^2 p}{\partial r^2}$
will benegative or zero over the entire line.
Consider now the Raychaudhuri equation,
\begin{equation}
\label{eq13}
2 \sigma^2 + \frac{1}{3} \theta ^2 +\frac{4\pi}{3} (\rho + 3p) = 
\dot v^\mu ;_{\mu} -\dot \theta   
\end{equation}
where 
$$\dot \theta = \theta,_{\mu} v^\mu = \frac{1}{\sqrt{g_{00}}}
\frac {\partial \theta}{\partial t}$$
and 
$\dot v^\mu;_\mu = \frac{1}{\sqrt {-g}} (\dot v^1 \sqrt {(-g_)}),_1$.

Now all the terms of the left hand side of (\ref{eq13}) are positive and the first term on the right
is zero or negative as we have just seen.
Hence $\dot \theta$ is negative at all points on the line and
$$ \int_{-\infty}^{+\infty} \frac {\partial \theta}{\partial t}dt > 0$$
which contradicts the condition of regularity that $\theta$ is
zero at both $t \rightarrow \pm \infty$.

In a similar manner it is easy to see that there is no other
zero of $\frac {\partial p}{\partial t}$ except at $t=0$.

\section{Expressions of the different variables in terms of
separated functions}

The results of the last section allow us to express $p$ in the
following form
$$p = \Sigma R_i T_i Q(r,t)$$
where $R'$s are linearly independent functions of $r$ having 
$\frac  {\partial R}{\partial r} 
=0$ at $r=0$ and vanishing at $ r \rightarrow \pm \infty$. 
$T'$s are similarly functions of $t$ alone. We have put in the summation
to take care of the possibility that there may be more than one function 
satisfying the imposed conditions.
(Indeed if $f(r)$ is one such function then any positive power of $f(r)$ 
will satisfy the same condition.) $Q(r,t)$ is any function 
involving both $r$ and $t$ (if possible). However, as there is no 
other zero of  
$\frac {\partial p}{\partial r}$
and $\frac {\partial p}{\partial t}$ except those
covered in $R'$s and $T'$s, the drivarive of $Q(r,t)$ with respect to
$r$ and $t$ cannot have any zero. Cosequently, if $Q$ is
to be bounded everywhere, it must be  a constant and can be absorbed in the  $R'$s and $T'$s. (It is assumed that the summation may be over infinite terms to 
include all possible  $R'$s and $T'$s.)

The functions $R$ and $T$ will be determined by the field equations
which are second order partial differential equations - hence there
can be at most two linearly independent $R$ functions and two $T$ functions.
So finally,
$$p = R_p T_p +\bar R_p \bar T_p \eqno{(14)}$$
and similarly
$$\rho = R_\rho T_\rho + \bar R _\rho \bar T_\rho \eqno{(14b)}$$
$$\ln g_{00} = R_0T_0  +\bar R_0 \bar T_0  \eqno{(14c)}$$
$$\ln h =
R_hT_h + 
\bar R_h\bar T_h + M(x^a) \eqno{(14d)}$$

where the unbarred and barred functions with the same subscript
are linearly independent.

As yet, we have not proved that $h$ is independent of the coordinates
$x^2,x^3$, hence we have added the term $M(x^a)$ in (14d).
No such term is required in the other three expressions as $p,\rho, \ln g_{00}$
have been shown to be functions of $r$ and $t$ only.

Putting in the expressions (14a,b,c) in equation (6), we get,
$$R'_pT_p + \bar R'_p \bar T_p = -\frac{1}{2}(R_pT_p + \bar R_p \bar T_p 
+R_\rho T_\rho +\bar R_\rho \bar T_\rho)(R'_0 T_0 +\bar R'_0 \bar T_0) \eqno{(15)} $$

where superscript primes indicate differentiation with respect to $r$.
 
Using now the non-linear dependence of the barred and unbarred functions, we get from equation (15),

$$\ln g_{00} = R_0 + T_0   \eqno{(16a)}$$
$$\rho = a R_p T_p +  b \bar R _p \bar T_p \eqno{(16b)}$$
$$p = R_pT_p +\bar R_p \bar T_p \eqno{(16c)}$$

where $a$ and $b$ are constants and 
$R_0, T_0$ are different from those in (14c). Also,
$$\ln \bar R_p = \frac{1+b}{1+a} \ln R_p$$
and $$ R_0 = -\frac{2}{1+a} \ln R_p = -\frac{2}{1+b}ln \bar R_p$$.

Similarly plugging in (14d) in equation (7), we get,

$$ \ln |h| = R_h + T_h + M(x^a) \eqno {(17a)}$$  
$$\ln \bar T_p = \frac{1+b}{1-a} \frac{a}{b} \ln T_p \eqno{(17b)}$$
$$ T_h = -\frac{2}{1+a} \ln T_p \eqno{(17c)}$$.

We have already noted that $g_{11}$ and $\gamma$ also
blow up at the infinities of $r$ and $t$. Hence they also must be expressible
in form similar to the above. Keeping in mind the relation $|h| = |\gamma ||g_{11}|$,
we can write,
$$\ln |g_{11}| = R_1 + T_1 + N(x^a) \eqno{(18a)}$$
$$ \ln |\gamma | = R_\gamma + T_\gamma + P(x^a) \eqno{(18b)}.$$
Note that we can make transformations $\bar t =f(t)$ and
$\bar r = \phi (r)$ to have
$$\ln |g_{11}|  - \ln g_{00} =  N(x^a) $$
or, $g_{11}/g_{00} =$  a function of $x^a$. 

\section{Proof of the existence of $G_2$}

We shall proceed in several steps. The first step is to prove that in the 
solutions that we are seeking, shear must be non-vanishing.

For this it is necessary to use the field equation 
$$R_{0i} = -8 \pi (T-T_{0i}) \eqno{(19)}$$

The above equation for a perfect fluid in irrotational motion
can be written in the form
$$\sigma^\alpha_{i;\alpha} - 
\sigma^{l}_i\dot v_{l} = \frac{2}{3} \theta ,_i \eqno{(20)}$$
In our present coordinate system this becomes
$$\sigma^k_{i|k}  = \frac{2}{3} \theta ,_i \eqno{(21)}$$
where $|k$ indicates covariant differentiation with the three
space metric. If now $\sigma_i^k$ vanishes, $\theta $ will be spatially 
uniform and then either the situation is static ($\theta = 0$)
or the solution will have singularities.

If the three non-zero values of shear eigenvalues are $S_1,S_2 $ and $S_3$,
we get
$$S_1 = [\frac{1}{2} \frac {\partial }{\partial t}(\ln |g_{11}| - \frac{1}{6}
\frac {\partial}{\partial t}(\ln h)] \frac {1}{\sqrt{g_{00}}} \eqno{(22)}$$
i.e., it is a function of $r$ and $t$ only.

In the next step we show that 
$\sigma^2$ and $\dot v;_{\mu}^{\mu}$ are independent of the coordinates $x^a$. Looking back at equation (13),
we see that in view of (16) and (17) all the terms except 
$2\sigma^2$ and $\dot v;_{\mu}^{\mu}$ are functions of $r$ and $t$ only. 
Hence we must have,
$$ (2\sigma^2),_a  = (\dot v;_{\mu}^{\mu}),_a$$
But 
$$\dot v;_{\mu}^{\mu} = \frac{1}{\sqrt{-g}} (\dot v_1 g^{11} \sqrt{-g}),_1$$
$$= -\frac{e^{N/2}}{g_{00}}[(\ln g_{00}),_{11} 
+ \frac{1}{2} [(\ln g_{00}),_1 (\ln \gamma ),_1]$$

Excepting $e^{N/2}$, $g_{00}$ and the term within the braces
are  functions of $r$ and $t$ only and these have no zero for finite 
values of $r$  and $t$. Thus,
$$(2\sigma^2),_a = (e^{N/2}),_a \times {\rm{ ~~ a ~~non-vanishing ~~expression~~
depending~~ on~~}} r {\rm{~~and~~}} t.$$
Consequently, if $N$ is not a constant $(2\sigma^2),_a$ or $(\dot v;_{\mu}^{\mu},_a$ will have 
no zero in the finite region of $r$ and $t$. But as $\sigma^2$ vanishes
at $t=0$ for all values of $r$ and $x^a$ and  $(2\sigma^2),_a$ or $(\dot v;_{\mu}^{\mu},_a$ also vanish there,
and it follows that $N$ is a constant and hence
$$S_{1,a} = S_{2,a} = S_{3,a}= 0 \eqno{(23)}$$
$$ g_{11,a}=0$$
We now come to the last  step of our proof of existence 
of $G_2$. For this we introduce three orthonormal vectors
$\lambda_{1|}^i, \lambda_{2|}^i, \lambda_{3|}^i$,  the three
unit eigenvectors of the shear tensor. Using the following formulas given
in Eisenhart's book \cite{eisen}
$$\sigma_{ij;k} \lambda_{h|}^i \lambda_{l|}^j \lambda_{m|}^k
=(S_h -S_l) \gamma_{hlm} ~~~~~(h \neq l)$$
$$\sigma_{ij;k} \lambda_{h|}^i \lambda_{h|}^j \lambda_{l|}^k
=\lambda_{l|}^k \frac {\partial S_h}{\partial x^k}$$
we get from (21),
$$\frac{2}{3} \theta ,_1 = [S_{1,1} + \frac{1}{\lambda _{|1}^1} (S_1 -S_2) \gamma_{122} 
+\frac{1}{\lambda_{|1}^1}(S_1-S_3)\gamma_{133}] \eqno{(24)}$$
$$0 = [S_{2,2} + \frac{1}{\lambda _{|2}^2} (S_2 -S_3) \gamma_{233} 
+\frac{1}{\lambda_{|2}^2}(S_2-S_1)\gamma_{211}] \eqno{(25)}$$
$$0= [S_{3,3} + \frac{1}{\lambda _{|3}^3} (S_3 -S_2) \gamma_{322} 
+\frac{1}{\lambda_{|3}^3}(S_3-S_1)\gamma_{311}] \eqno{(26)}$$
where $\gamma'$s are Ricci relation coefficients. 
As $g_{11,a} = 0$, $\lambda_{1|}^i$ is geodetic in the three
space. Hence $\gamma_{211} =\gamma_{311} =0.$ Now from
(25) and (26) as $S_{i,a} = 0$ we get, if
$S_2 \neq S_3$
$$\gamma_{233} = \gamma_{322} =0  \eqno{(27)}$$
If $S_2 = S_3$, then the existence of $G_2$ can be readily
proved but as will be apparent from our later discussion,
this case will lead to the vanishing of shear and
hence to singular solutions. From (27), owing to the antisymmetry $\gamma_{abc}
=-\gamma_{bac}$, it follows that all the $\gamma 's$ involving the
subscripts 2 and 3 vanish and consequently
the two spaces spanned by $x^2$ and $x^3$ are all flat;
i.e., we may choose coordinates such that $g_{ab,c} =0$. Thus all the
metric tensor components are independent of $x^a$ - the space time
admits translations along $x^2$ and $x^3$.

\section {Cylindrical Symmetry}

From the field equations, one can deduce in case
of irrotational motion of a perfect fluid (cf. Ellis \cite{ellis}).
$$^*R = -\frac{2}{3} \theta^2 + 2 \sigma^2 +16 \pi\rho \eqno{(28)}$$
where $^*R$ is the Ricci scalar for the three space
$t =$ const. Now by direct calculation,
$$^*R^1_1 = - \frac{1}{\sqrt{-g_{11}}} \frac {\partial \Theta}{\partial r}
-\frac{1}{2} \Theta ^2 -2 \Sigma^2 \eqno{(29)}$$
where $\Theta$ and $\Sigma$ are the 'expansion' and 'shear scalar'
for the unit space-like vector $\xi^i = \frac{1}{\sqrt{-g_{11}}} \delta ^i_1$, i.e.,
$$ \Theta = \xi^i_{|i} = \frac{1}{\sqrt{-g_{11}}} (\ln \sqrt{\gamma}),_1$$
$$2 \Sigma ^2 = \Sigma _{ik} \Sigma^{ik}$$
$$ \Sigma _{ik} = \frac{1}{2} (\xi_{i|k} +(\xi_{k|i}) - \frac{1}{2} (g_{ik} + (\xi_i  \xi_k) \xi^m_{|m}$$
where $|i$ indicates covariant differentiation with  the three
spaces metric. The reader may note the similarity between
equation (29) and the Raychaudhuri equation in the 4-dimensional
space time. Again with $G_2$,
$$^*R = ^*R_1^1 - \frac{1}{\sqrt{-g_{11}}}  \frac {\partial \Theta}{\partial r}
- \Theta^2 \eqno{(30)}$$
Combining (28), (29) and (30), one gets,
$$ -\frac{2}{\sqrt{-g_{11}}}  \frac {\partial \Theta}{\partial r} -\frac{3}{2} \Theta^2 =    - \frac {2}{3} \theta^2 + 2 \sigma^2 + 
16 \pi \rho + 2 \Sigma^2 \eqno{(31)}$$ 
For the three space at $t = 0$, $\theta^2 = \sigma^2 = 0$ and we get from 
(31)
$$ -(\frac{2}{\sqrt{-g_{11}}}  \frac {\partial \Theta}{\partial r} + \frac{3}{2} \Theta^2) > 0 \eqno{(32)}$$
The inequality 
(32) shows that $\Theta$ would blow up leading to the collapse of the two-space
(i.e., $\gamma = 0$). This will not mean a singularity iff $x^2$ or $x^3$ or both be angle
coordinate so that $\gamma$ will naturally vanish at $r=0$. However both cannot be angle
coordinates for then $\Sigma^2$ would vanish at
$r = 0$ and (31) would make $\rho$ blow up at that point.
With one only of $x^2,x^3$ being angle coordinate both the left
hand side as well as $\Sigma^2$ would blow up
leaving a finite difference for $\rho$.

Thus one of the translations is along an angle coordinate indicating that we 
have cylindrical symmetry.

\section{Relation between different time functions}

Let us identify $x^2$ as the angle coordinate. Then the
condition of elementary flatness gives $S_1 = S_2$. Plugging
in this result in (24), we get
$$S_{1,1}  + 3\frac {S_1}{2} (\ln g_{33}),_1 = \frac {2}{3} \theta,_1$$
or
$$ \frac {1}{2} {S_1} (\ln g_{00}),_1 -  3\frac {S_1}{2} (\ln g_{33}),_1
= \frac{1}{3} \theta  (\ln g_{00}),_1$$
so that
 $$ \frac {(\ln g_{33}),_1}{  (\ln g_{00}),_1} = \frac {2}{3} \frac 
{\frac{(S_1}{2} - \frac{\theta}{3})}{S_1} \eqno{(33)}$$
From what we have proved it is clear that the left hand side of
(33) is a function of $r$ alone while
the right hand side a function of $t$ only.
Hence both sides must be constant. So that we have
$$S_2 = S_1 = n\theta; ~~~S_3 = -2n \theta \eqno{(34)}$$
where $n$ is a constant. Equation
(34) leads to the desired result that $T_1 = T_2 = T_0$, and $T_h$ are 
expressible as powers of a single time function. 
Equations (17b) and (17c) then show that $T_p$ and $\bar T_p$
are also of the same type. 
Hence we have deduced the assumption (c) of Ruiz and Senovilla.

At this stage it is clear that we have been able to deduce all the assumptions
made by Ruiz and Senovilla except the hypersurface orthonormality of the killing vectors.
This is as it should be for Mars has shown that non-singular solutions
do exist where the killing vectors are not hypersurface orthogonal. Also,
our demonstration of separatibility stops at that of
$\gamma$ and does not include the non-diagonal $g_{ab}$. This again is 
consistent with Mars's finding.

Thus we may claim to have achieved our objective namely that with our conditions
(i) to (v), the already known solutions given by Ruiz-Senovilla and Mars are
the only possible non-singular solutions.

\section{Concluding Remarks}

It is tempting to ask  whether the present work can be
extended 
to more general cases, i.e., whether one can establish specific conditions
about non-singular cosmological solutions when one or more of the conditions
of irrotational motion, perfect fluidity and $\frac {\partial p}{\partial \rho}
> 0$ is given up. Already there exist
quite a number of imperfect fluid non-singular solutions but they vary widely
in their characteristics. Thus besides cylindrically symmetric 
solutions, there are also spherically symmetric solutions \cite{dadhich3}
and even an oscillatory solution \cite{dadhich4}. In fact for imperfect
fluids, the divergence relations 
$T_{\mu;\nu}^\nu = 0$ written out explicitly are much more complicated than (6) and (7)
and do not seem amenable for simple general analysis. again if rotation be present, non-diagonal terms $g_{0i}$ appear in the metric tensor and further
calculation becomes much involved. So far as the condition $\frac {\partial p}{\partial \rho}
> 0$
is concerned, the present author believes that it can be done away
with although right now he is not in  a position to show that.

Acknowledgements: 

The author's thanks are due to the members of the Relativity and Cosmology 
Center, Jadavpur University and N. Dadhich of Inter University center
for Astronomy and Astrophysics, Pune (India), for helpful
discussions.

\end{multicols}

\begin{references}
\bibitem{seno} J. M. M. Senovilla, Phys. Rev. Lett. {\bf 64} ,2219 (1990).
\bibitem{rseno} E. Ruiz and J. M. M. Senovilla, Phys. Rev.  {\bf D45}, 
1995 (1992).
\bibitem{mars} M. Mars,  Phys. Rev.  {\bf D51}, R3989 (1995).
\bibitem{ear} J. Earman,{\it The expanding World of General 
Relativity}, Einstein studies {\bf 7} ed. H. Goenner et al, Boston, Barkhauser 
Verlag (1998).
\bibitem{dadhich}N. Dadhich, L. K. Patel and R.Tikekar, Pramana - Journal 
of Physics, {\bf 44}, 303 (1995).
\bibitem{dadhich2} {\it Inhomogeneous Cosmological Models}, eds. 
A. Molina and J. M. M. Senovilla, World Scientific, 
Singapore (1999), Pp 63-74.
\bibitem{seno2} J. M. M. Senovilla, Phys. Rev. D {\bf 53}, 1799 (1996).
\bibitem{akr1} A. K. Raychaudhuri, Phys. Rev. Lett. {\bf 80}, 654 (1998).
\bibitem{akr2} A. K. Raychaudhuri, Mod. Phys.  Lett  {\bf  A15}, 391 (2000).
\bibitem{eisen} L. P. Eisenhart, {\it Riemannian Geometry}, Princeton 
university Press, Princeton (1950).
\bibitem{ellis} {\it General Relativity and Cosmology},
Proceedings of the International School of Physics, 'Enrico Fermi' Course
XLVII, ed. R. K. Sachs, Academic Press (1991)
\bibitem{dadhich3}N. Dadhich,  J. 
Astrophys. Astro  {\bf 18}, 343 (1997).
\bibitem{dadhich4}N. Dadhich and A. K. Raychaudhuri,  Mod. Phys. Lett. {\bf A14},
2135 (1999). 

 
\end{references}
\end{document}